\documentclass[twocolumn,showpacs,aps,pra]{revtex4-1}
\usepackage{graphicx}
\usepackage{dcolumn}
\usepackage{bm}
\usepackage{amsthm}
\usepackage{amsmath}
\usepackage{amssymb}
\usepackage{color}
\usepackage{epsfig}
\usepackage{longtable}
\usepackage {booktabs}
\usepackage{comment}
\usepackage{tabularx}
\usepackage{mathtools}
\usepackage{placeins}

\definecolor{Red}{rgb}{1,0,0}

\def\ket#1{| #1 \rangle}

\begin{document}

\title{Optimizing the Frequency of Quantum Error Correction using the [[7,1,3]] Steane Code} 

\author{Ali Abu-Nada$^1$, Ben Fortescue$^1$,  Mark Byrd$^{1,2}$}

\affiliation{$^1$Physics Department, Southern Illinois University, Carbondale, IL 62901, USA}
\affiliation{$^2$Computer Science Department, Southern Illinois University, Carbondale, IL 62901, USA} 

\date{\today}

\begin{abstract}
A common assumption in analyses of error thresholds and quantum computing in general is that one applies fault-tolerant
quantum error correction (FTQEC) after every gate.  This, however, is known not to always be optimal if the FTQEC procedure
itself can introduce errors.  We investigate the effect of varying the number of logical gates between FTQEC operations, and
in particular the case where failure of a postselection condition in FTQEC may cause FTQEC to be skipped with high
probability. 

By using a simplified model of errors induced in FTQEC, we derive an expression
for the logical error rate as a function of error-correction frequency, and show that in this model the optimal frequency is relatively insensitive
to postselection failure probability for a large range of such probabilities.  We compare the model to data derived from Monte Carlo simulation
for the $[[7,1,3]]$ Steane code.

\end{abstract}


\maketitle

\section{\label{sec:level1} Introduction}

Quantum error correcting codes (QECC) are often required to correct errors due to imperfectly implemented quantum operations during quantum information processing and also to simply store quantum information \cite{nielsen,shor95,steane96,Calderbank96}.  For a quantum error correcting code to be effective, the probability for an error to occur on any given gate needs to be below the threshold value for the code \cite{aharonov08}.
However, even if the error rate is below the threshold, the resources required for quantum error correction can greatly exceed those resources required for the quantum information processing alone.  

QECCs encode logical qubits into multiple physical qubits; 
additional quantum error correction (QEC) operations are regularly performed to diagnose and repair errors that may have arisen.
(In this paper we use QEC to denote the operations required to obtain an error syndrome from data and correct any detected errors, although
in the examples we consider the latter does not require any additional physical operations).

A common model for such an error-corrected computation is to apply a QEC immediately 
after each logical gate in order to correct the physical
errors introduced by these gates before they accrue into more serious logical errors. However, QEC operations will, in general,
also introduce errors with some non-zero probability since they also use imperfect physical gates.  These are often the same gates used to perform logical operations.  
In addition to the threshold needing to be satisfied, for QEC operations to successfully suppress errors enough to allow large-scale computation, the QEC as well as logical gates 
must be implemented in a fault-tolerant way
\cite{preskill98}, $i.e.$, in a way such that a single faulty quantum gate cannot lead to multiple errors on the data. 
Fault-tolerant constructions for QEC operations are often dependent on post-selection;
in particular, they commonly use ancillary states to diagnose errors, which effectively carry away entropy from the data.
Since they interact with the data, these states must be prepared so that they do not spread multiple
errors to the data. Often, however, non-fault-tolerant
circuits must be used for the initial ancilla preparation, and fault-tolerance is instead enforced by post-selection
(ancilla verification) in which only those ancillas which satisfy some measurement outcome after being created
are subsequently used to interact with the data \cite{shor95,steane96}.

This model of fault-tolerant QEC, in which ancillas are created until one passes post-selection, can be difficult to implement.  The data can accumulate
additional errors either waiting for a ``good'' ancilla to be created (if created sequentially) or to be moved into place (if created in parallel).
And the nondeterministic delays involved make synchronization of the data with other data blocks difficult.  An alternative, which preserves synchronization, is to simply to allow for 
a fixed number of postselection attempts, and to carry on with the computation (skipping the QEC) if these are unsuccessful.  The obvious disadvantage is that skipped QEC 
operations allow errors to accumulate from sequential logical gates, but if the skipping probability and gate error probabilities are sufficiently low this may
 still be the optimal solution.  Another alternative is go through with the error correction without post-selection and instead measure to detect any errors
 on the ancilla and correct for those as well as any error already present.  This is called ancilla decoding and was originally shown to be advantageous in the regime of
slow, noisy measurements \cite{div07}, but has recently been considered in other contexts \cite{ali14}.

A closely related situation is where QEC operations are intentionally not applied after every gate, but only after a certain number of gates.  This can reduce
the overall logical error if the errors introduced by the QEC operation are larger than those introduced by the gates.
Recently, Weinstein has provided a specific relation between  the  fidelity  and
physical error rates for performing QECs after different numbers of gates in the $[[7,1,3]]$ Steane code \cite{yaa13}. Here, it is  shown that the overall error rate
is not significantly increased by performing QEC after two gates rather than after every gate, and
that the inherently noisy correction process will sometimes introduce more error into the system
when applied too frequently \cite{yaa13,yaa14}. 

In this paper, we consider the optimal frequency to apply postselection-dependent QEC operations, in the case where postselection failure
results in a skipped QEC.  We consider a model where a logical data qubit undergoes operations from $N$ logical gates, with  $m$ gates in between each QEC, using the well-known [[7,1,3]] Steane code \cite{steane96} and the Steane ancilla
technique for the latter.  We determine the logical error rate $P_{L}$ for the data after undergoing these $\left(N/m\right)$ ``blocks'' of gates and QEC operations,
as a function of $m$ and the underlying physical gate error $\epsilon$, and then minimize as a function of $m$.

In section \ref{sec:level2} we explain the model and the analytical derivation of the formula given the model.  In section \ref{sec:level3} we compare the predicted
error rate of the model to that obtained via Monte Carlo simulation.

\section{\label{sec:level2} Derivation of the mathematical formula for $P_L$}
\subsection{Logical gate and QEC model}
In general, the dependence of the logical error rate on the physical error rate of the underlying gates depends on the circuits used to implement gates and QEC.
These circuits can be complex, and highly dependent on the chosen code.  While
our analysis is based on the Steane code, we wished to use a model that could be readily adapted to other codes.  We thus produced a semi-abstracted
model based on the Steane code with Steane ancillas.

Errors may be introduced into the logical qubit in two ways, from the logical gates and the `` noisy`` QEC itself (in this analysis we
do not treat errors from movement or hold operations as a separate category, they may incorporated into the above
categories if desired).  We model a noisy physical gate as performing the desired operation followed by, with probability $\epsilon_{g}$,
an error operation.  We treat logical gates as transverse, that is, simply consisting of a single physical gate applied to each qubit.

\begin{figure}
\includegraphics[scale=0.5]{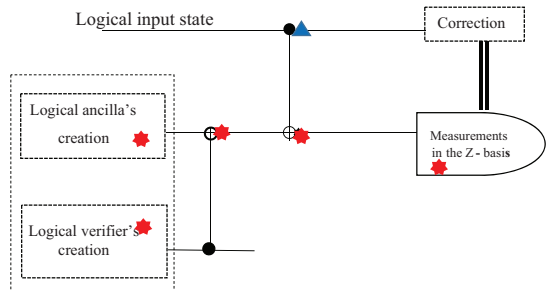}
\caption{\label{fig:diagram} A schematic diagram for a ``noisy`` QEC for correction of $X$ errors, where the  star represents  the location where a syndrome error
 might occur, and the triangle a location where a correction error  might occur. The Steane ancilla state must be prepared in the $\ket{ +_{L} }$ and
the measurements performed in the Z-basis.}
\end{figure}

Our model of the QEC operation is more approximate.  We divide errors induced by QEC into four separate parts
(as shown in Figure \ref{fig:diagram}) with the following probabilities:
\begin {itemize}
 \item ``Correction errors'', with probability $\epsilon_{c}$ per qubit, are due to physical errors in those gates directly applied to the data as part of the QEC process.  In the Steane code
and many others, this is limited to those two-qubit gates used to interact the data with the ancillas (since the corrections themselves can be done using the ``Pauli frame'' \cite{steane03}, without
the use of physical gates).  We defined correction errors as being those which affect the data {\it only}, not the QEC ancilla.  Hence two-qubit gate failures which lead to
errors on both outputs (both data and ancilla qubits) are excluded.
\item The remaining errors are those which affect the data through causing an incorrect syndrome measurement, and thus an incorrect correction operation.  We first define
``syndrome errors'', with probability $\epsilon_{s}$ per qubit, as those errors where an error on the QEC ancilla (or its measurement) {\it only} (with no errors directly
applied to the data) cause a data qubit to be wrongly ``corrected'', in addition to any errors already present.
\item ``Omission errors'', with probability $\epsilon_{o}$ per qubit, represent the special class of erroneous syndromes which coincidentally combine with existing data errors (where present) to (wrongly) return
a syndrome indicating no errors.  Thus if the data is initially without error, an omission error in the QEC will lead to an error on the data, but if a single qubit error is present
on the data prior to the QEC, an omission error will simply lead to this error remaining uncorrected.
\item Finally, ``double errors'', with probability $\epsilon_{d}$ per qubit, are those errors where a single gate failure in the CNOT joining the data to the ancilla lead to $X$
errors on both the source and the target (thus we only need consider this class of errors when modeling source and target errors on two-qubit gates as
correlated).  This behaves differently to a standard syndrome error since if a double error occurs during syndrome measurement, is the only error occurring
during that syndrome measurement, and the data has no prior errors, no error will be added to the data from the overall QEC.  This is because an error on both
source and target qubits in the CNOT is equivalent to an error on the data which correctly propagates to the syndrome, and will therefore
be successfully corrected.  Conversely, since a double error does affect the data directly, if a double error on occurs on one qubit
and a syndrome (or double) error on another qubit as part of the same syndrome measurement,
this can lead to two errors on the data (while two standard syndrome errors could only lead to one error on the data).
\end{itemize}
We additionally consider the probability of an ``ancilla error'', with probability $\epsilon_{a}$ (not per qubit).  This is the case where a QEC operation (that is,
any elements of that operation that could either correct errors or produce errors on the data) is not performed (for example, due to failure in the postselection
process for the necessary ancilla).  Such events do not produce data errors themselves, but result in existing errors not being corrected when they should be.

Note that all of the above are functions of the physical gate error $\epsilon_g$, but the exact relationship depends on the circuits and QECC used,
thus we treat them as separate variables in our model.

\subsection{Logical error rate\label{sec-logical}}
The sequence of $(N/m)$ blocks may, in a distance-3 code, produce a logical error if  one or more of these  blocks create two or more physical errors on the logical qubit. 
Therefore, to estimate the logical error rate $P_{L}$, we must enumerate the ways in which these blocks  might create two or more physical errors on the logical qubit. This 
logical error might be created either within a single block, 
or across multiple blocks (if a QEC is either skipped due an ancilla error or fails due to a syndrome error.)  Thus a logical error will only
occur with probability second-order or higher in the various error probabilities.

We are particularly interested, however, in the regime where the ancilla errors are significantly larger than the other errors.  This can easily
be the case if ancilla creation circuits are complex.  While the postselection procedure means such large errors do not translate directly to logical errors,
they can result in multiple QEC operations being skipped, with a consequent increase in the logical error probability if other errors are present.

Considering now the structure of a block, this consists of $m$ gates, followed by a QEC operation consisting of a successful correction of any errors on the logical qubit,
unless an error occurs in the QEC.  Finally any correction errors are applied.
Thus, in the absence of QEC verification failures, the leading-order contributions to the logical error rate (where two errors occur resulting in a logical error) are very limited: two errors can occur within a block  or across two adjacent blocks, as shown
in detail in Tables \ref{table:errors-standard} to \ref{table:errors-gamma3}.

A verification failure permits additional error combinations: errors on two separate blocks between which a successful correction would ordinarily occur.
Note that additional verification failures permit, to second order, the same types of logical errors.
That is, the only second-order errors which $f$ skipped QECs additionally permit (beyond those which could occur regardless) are when the skipped QECs all occur sequentially, and the two errors in question are on the block containing the first skipped QEC and the block following the final skipped QEC.  For a sequence of $B\equiv N/m$ blocks, there are $B-f$ ways to have $f$ sequential skipped QECs.  Thus the additional logical error due to $f$ skipped QECs is weighted by a factor
\begin{align}
\gamma&\equiv\sum_{f=1}^{B-1}\epsilon_a^f(B-f)\\
&=\frac{B\epsilon_a(1-\epsilon_a)-\epsilon_a+\epsilon_a^{B+1}}{(1-\epsilon_a)^2}
\end{align}
as used in Table \ref{table:errors-gamma}.
Due to the definition of correction errors, in some cases the combined errors allowable due to a single skipped QEC may span 3 rather than 2 blocks, in which case the multiplying factor is
\begin{align}
\gamma_3&\equiv\sum_{f=1}^{B-1}\epsilon_a^f(B-f-1)\\
&=(\gamma-(B-1)\epsilon_a)/\epsilon_a\\
&=\frac{\epsilon_a}{(1-\epsilon_a)^2}\left(B(1-\epsilon_a)-2+\epsilon_a+\epsilon_a^{B-1}\right)
\end{align}
as used in Table \ref{table:errors-gamma3}.

In Tables \ref{table:errors-standard} to \ref{table:errors-gamma3}, we show the possible ways for a logical error to occur to second order in the various errors.
The final column represents the overall contribution to the logical error rate from all error combinations of that type. 
Note that rows represent a general class of errors (e.g. the first row in Table \ref{table:errors-standard} represents the contribution due to two gate errors within a single block) and the rows
involving factors of $\gamma$ and $\gamma_3$ represent classes of errors where $f$ intermediate QEC steps are skipped leading to an overall factor of $\gamma$ or $\gamma_3$.

Summing the terms, the general second-order formula  for $P_L$ is :
 \begin{widetext}
\begin{multline}
\label{formula}
P_L =42\Big[B\left(m\epsilon_g\left(\frac{m\epsilon_g}{2}+(1-\epsilon_a)(\epsilon_s+\epsilon_d)\right)
+(1-\epsilon_a)\left(\epsilon_c\left(\epsilon_s+\epsilon_o+\frac{\epsilon_c}{2}\right)+\epsilon_d\left(\epsilon_s+\frac{\epsilon_d}{2}\right)\right)\right)\\
+(B-1+\gamma_3)(1-\epsilon_a)(\epsilon_c+\epsilon_s+\epsilon_o)(m\epsilon_g+(1-\epsilon_a)(\epsilon_s+\epsilon_d))
+\gamma m\epsilon_g\left(m\epsilon_g+(1-\epsilon_a)(\epsilon_s+\epsilon_d)\right)\Big]
\end{multline}
\end{widetext}

\subsection{Minimizing $P_L(m)$}
$P_L$ is a discrete function of $m$.  In the limit of many blocks ($B\to\infty$) we can neglect express $P_L$ as a function 

\begin{equation}
P_{L_p}(m)\simeq dm^{-1}+c_0+c_1m\label{eq-plm}
\end{equation}
where
\begin{align}
d&=42B(1-\epsilon_a)\left[\begin{multlined}\epsilon_c\left(\epsilon_s+\epsilon_o+\frac{\epsilon_c}{2}\right)+\epsilon_d\left(\epsilon_s+\frac{\epsilon_d}{2}\right)\\
+(\epsilon_s+\epsilon_d)(\epsilon_c+\epsilon_s+\epsilon_o)\end{multlined}\right]\\
c_0&=42B\epsilon_g(\epsilon_c+2\epsilon_s+\epsilon_o+\epsilon_d)\\
c_1&=42B{\epsilon_g}^2\left(\frac{1}{1-\epsilon_a}-\frac{1}{2}\right)
\end{align}

$m_{min}$ satisfies
\begin{align*}
P_L(m_{min})&<P_L(m_{min}-1)\\
P_L(m_{min})&<P_L(m_{min}+1),
\end{align*}
thus from \ref{eq-plm} we have
\begin{align*}
m_{min}(m_{min}+1)c_1-d&>0\\
m_{min}(m_{min}-1)c_1-d&>0
\end{align*}
and hence (since $m_{min}$ is positive)
\begin{align*}
m_{min}&>\sqrt{\frac{1}{4}+\frac{d}{c_1}}-\frac{1}{2}\\
m_{min}&<\sqrt{\frac{1}{4}+\frac{d}{c_1}}+\frac{1}{2}
\end{align*}

and so $m_{min}$ is the unique integer satisfying
\begin{equation}
\sqrt{\frac{1}{4}+\frac{d}{c_1}}-\frac{1}{2}<m_{min}<\sqrt{\frac{1}{4}+\frac{d}{c_1}}+\frac{1}{2}\label{eq-mmin}.
\end{equation}
Thus the dependence of $P_L$ on $m$ is determined by the variable
\begin{widetext}
\begin{equation}
\frac{d}{c_1}=\frac{2(1-\epsilon_a)^2[\epsilon_c\left(\epsilon_s+\epsilon_o+\frac{\epsilon_c}{2}\right)+
\epsilon_d\left(\epsilon_s+\frac{\epsilon_d}{2}\right)+(\epsilon_s+\epsilon_d)(\epsilon_c+\epsilon_s+\epsilon_o)]}{\epsilon_g^2(1+\epsilon_a)}.
\end{equation}
\end{widetext}

\section{\label{sec:level3} Monte Carlo Simulation of $P_L$}
As discussed above, our analytical formula for the logical error simplifies the description of QEC errors (in general a function of complex ancilla circuits) to the variables
$\epsilon_{s,o,c,d}$, which we assume to take the same set of values for every qubit.  In order to check the accuracy of this approximation, we performed  Monte Carlo simulations
of the complete QEC for the [[7,1,3]] Steane code  with  the Steane ancilla technique, using QASM-P, simulation software based on QASM \cite{cross06},
in order to compare the logical error rates obtained with those predicted.

Initially, all gates were simulated using the stochastic error model for depolarizing noise \cite{ali14}.  In this case, we considered bit-flip ($X$) errors only on the data qubits (phase-flip ($Z$) errors may be dealt with independently in the [[7,1,3]] code, and in our chosen noise model, will occur at equal rates).  We used $N=1000$ with varying block sizes $m\in\{1,2,4,5,8,10,20,25,100\}$, thus $B$ varied between 1000 and 10.

Ancilla verification is performed as usual for $X$ correction using the Steane technique in the $[[7,1,3]]$ code: an ancilla interacts with the data as the control for a transversal CNOT gate, and needs to be prepared and verified so that a single gate failure will not cause the ancilla to have multiple errors which would be transferred to the data.  Usually, then, ancilla verification failure rates are, like the other errors, a function of the gate error rates.  However, in order to vary $\epsilon_a$ independently of other errors to verify the formula, our simulation artificially determines beforehand whether a verification failure error will occur.  If so, the QEC is skipped.  If not, the preparation and verification is repeated until passed.  Thus verified ancillas have the correct error statistics for a given gate error, but failure occurs with the chosen probability $\epsilon_a$.

To determine the logical error probability $P_L$, the data is prepared, 
without error, in a logical $\ket{\overline{0}}$ state.  Then, as described above, a series of blocks of $m$ transversal logical gates (since we only wish to simulate errors, these are simply wait operations which in the absence of errors do not change the qubits' state), followed by one QEC operation per block, are applied, for a total of $N$ logical
 gates and $B=N/m$ blocks and (attempted) QECs.   Finally,  the data is checked for logical $X$ errors.  Each simulation (for a given choice of variable values $\epsilon_g$ and $\epsilon_a$ has $10^{6} $ runs).  From the Monte Carlo simulation we therefore obtain $P_{L}$ as a function of the underlying physical error rates.

\subsection{Numerical estimation of  $\epsilon_{s}$ and $\epsilon_{o}$}
By our definition, the only source of correction errors is the CNOT gate interacting the data with the ancilla.  Such errors occur only when the gate failure leads to an $X$ error on the CNOT source (the data) but not the CNOT target.  Similarly double
errors occur when a CNOT failure leads to X errors on both outputs.
In our depolarizing error model single qubit gates undergo $X$, $Y$ or $Z$ errors with equal probability $\epsilon/3$, and two-qubit gates
undergo the 15 possible two-qubit errors $(X\otimes I, Y\otimes I\ldots Z\otimes Z)$ with equal probability $\epsilon/15$.  Since our analysis only considers bit errors (introduced by either $X$ or $Y$ operators),
we have a single-qubit gate bit error probability of $\epsilon_g=2\epsilon/3$.  Thus the probability of a CNOT source-only error, which comes from the operations $X\otimes I, X\otimes Z, Y \otimes I, Y \otimes Z$ is
$\epsilon_c=4\epsilon/15=2\epsilon_g/5$.  Likewise the probability of of a double error comes from $X\otimes X, X\otimes Y, Y \otimes X, Y \otimes Y$ and hence $\epsilon_d=2\epsilon_g/5$

$\epsilon_{s}$ and $\epsilon_{o}$ were determined directly from simulation.  By our definition, a syndrome error or double error in a QEC will, for an input logical qubit containing one error,
add a second error, leading to an overall logical error, and are the only first-order QEC errors which do this.  Thus
to estimate error rate $\epsilon_{s}+\epsilon_d$, the data was first prepared  in a logical eigenstate, with an error on one of the seven qubits.
The QEC procedure for the [[7,1,3]] Steane code  with  the Steane ancilla technique was then performed using the stochastic error model for depolarizing noise \cite{ali14}. Finally the logical qubit was checked for logical errors to determine the error rate.

Similarly, if the input data has a single logical error entering and leaving the QEC, this will be due to either a correction or omission error. 
Hence to estimate the QEC physical error rate $\epsilon_{c}+\epsilon_o$, we prepare the input logical qubit with an error 
on one of the seven corresponding physical qubits, then we perform  the same QEC simulation procedure as before,
but determine the rate based on events when the output has a single error (rather than two).  In both cases we varied the input error over all 7 qubits and took the mean resultant $P_L$.

We performed  the simulation with  a variety of numerical  values for $ \epsilon_{g}$ ($10^{-5} \sim 10^{-3}$). For each different value
of  $ \epsilon_{g}$, we determine the numerical values of $ \epsilon_{s}$ and $ \epsilon_{c}+\epsilon_o$ The relationship is fitted to a linear equation to determine the coefficients
for $ \epsilon_{s}$ vs. $\epsilon_g$ and $ \epsilon_{c}+\epsilon_o$ vs.  $ \epsilon_{g}$, which were as follows:

\begin{align}
\epsilon_{s}+\epsilon_d&= 3.85 \epsilon_{g}\label{e_s}\\
\epsilon_{c}+\epsilon_o &= 1.01 \epsilon_{g}\label{e_c}\\
\Rightarrow \epsilon_s&=3.45\epsilon_g\\
\epsilon_o&=0.61\epsilon_g
\end{align}

\section{\label{selective} Results and discussions}

\begin{figure*}
  \centering
  \includegraphics[width=0.7\textwidth]{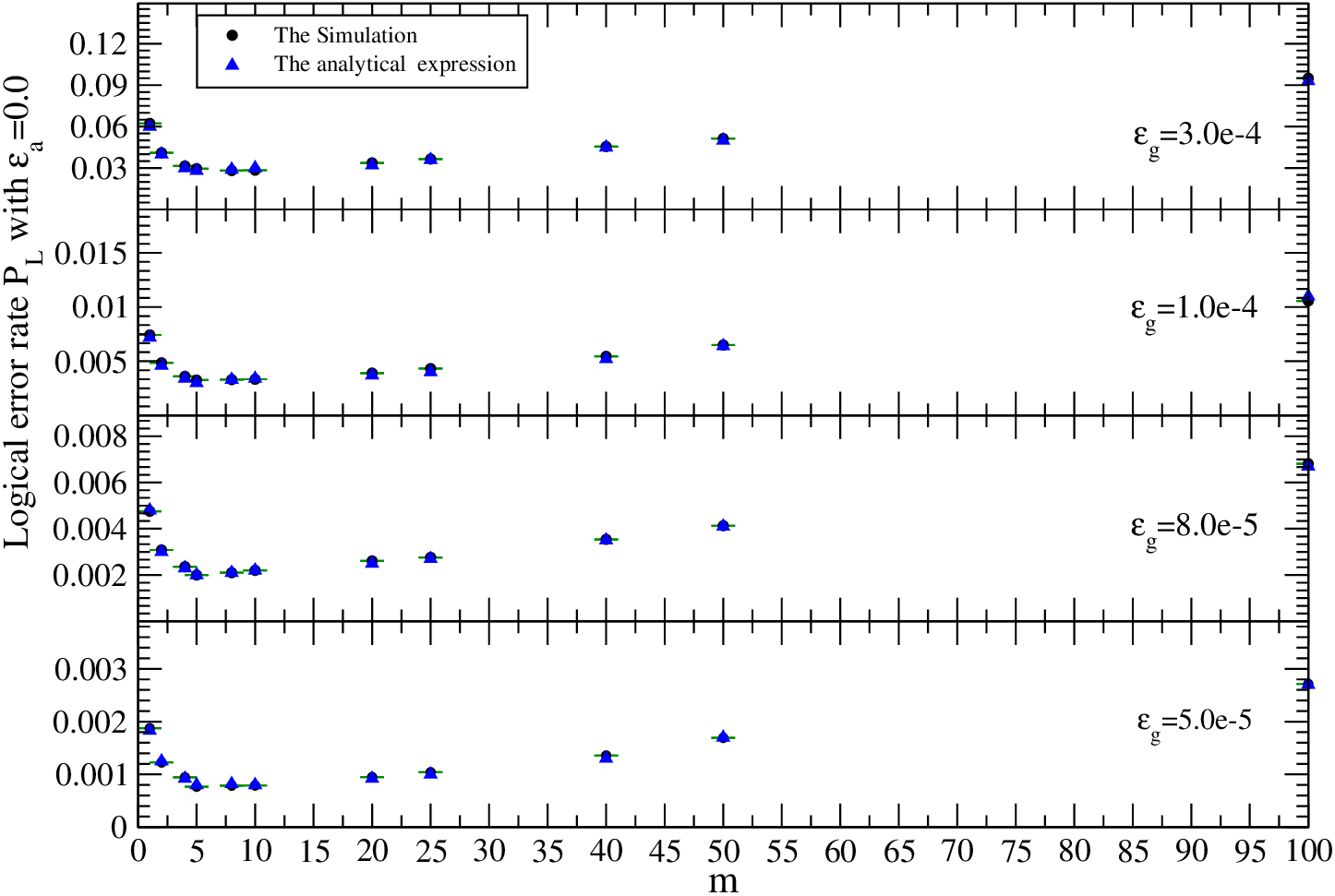}\\
  \caption{ $P_{L}$ vs $m$ for $\epsilon_{g}$= $5.0 \times 10^{-5}$ ,$8.0 \times 10^{-5}$,  $1.0 \times 10^{-4}$ ,  $3.0 \times 10^{-4}$,
where $\epsilon_{a}$ for this case is zero.The triangular blue  points are the numerical values of $P_L$ as given by the equation \eqref{formula}
. The circular black points are the numerical simulation values of $P_L$.
  \label{fig:e_a_zero}}
\end{figure*}

\begin{figure*}
  \centering
  \includegraphics[width=0.7\textwidth]{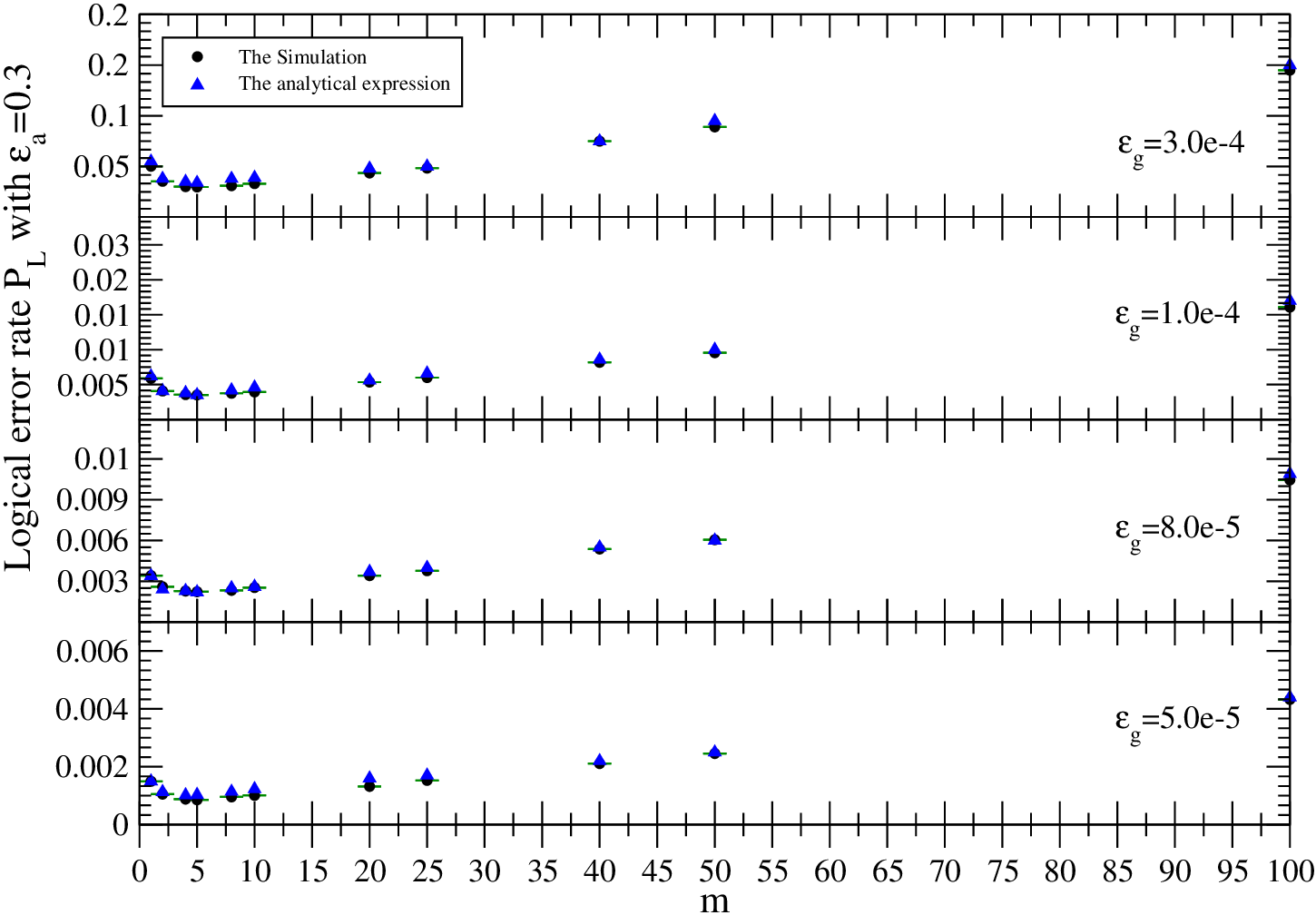}\\
  
\caption{ $P_{L}$ vs $m$ for $\epsilon_{g}$= $5.0 \times 10^{-5}$ ,$8.0 \times 10^{-5}$,  $1.0 \times 10^{-4}$ ,  $3.0 \times 10^{-4}$,
where $\epsilon_{a}$=0.3.}
  \label{fig:e_a_0.3}
\end{figure*}

\begin{figure*}
  \centering
  \includegraphics[width=0.7\textwidth]{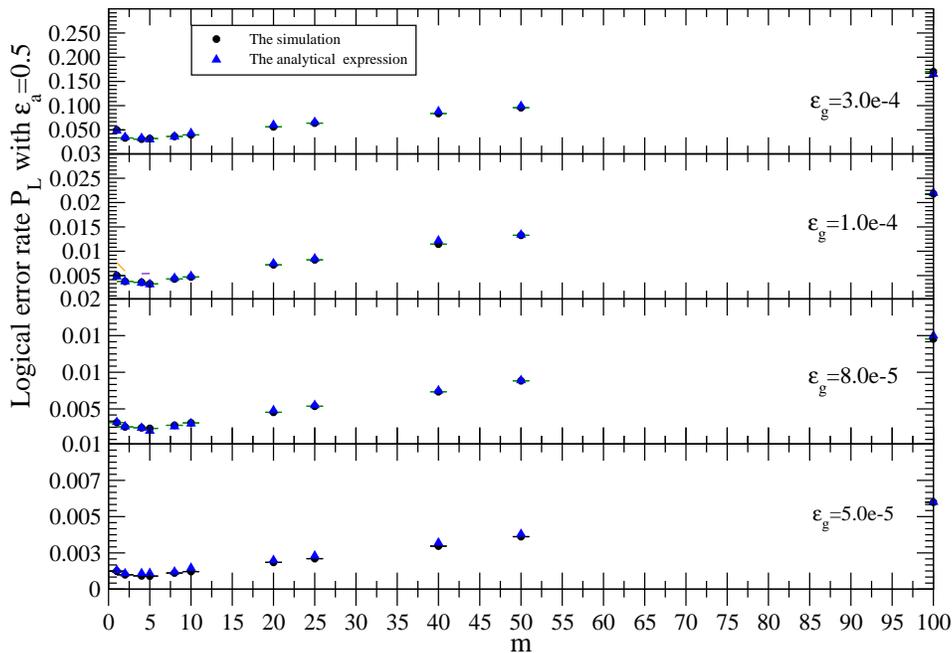}\\
  \caption{ $P_{L}$ vs $m$ for $\epsilon_{g}= 5.0 \times 10^{-5}$ ,$8.0 \times 10^{-5}$,  $1.0 \times 10^{-4}$ ,  $3.0 \times 10^{-4}$,
where $\epsilon_{a}$ =0.5.}
  \label{fig:e_a_0.5}
\end{figure*}

\begin{figure}
  \centering
  \includegraphics[width=0.45\textwidth]{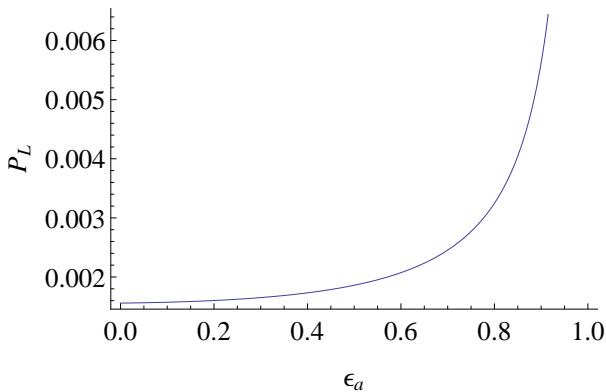}\\
  \caption{$P_{L}$ vs $\epsilon_a$ for $\epsilon_{g}$= $5.0 \times 10^{-5}$, $m=5$, using equation \eqref{formula} only}
  \label{fig:pl-ea}
\end{figure}

\begin{figure}
  \centering
  \includegraphics[width=0.45\textwidth]{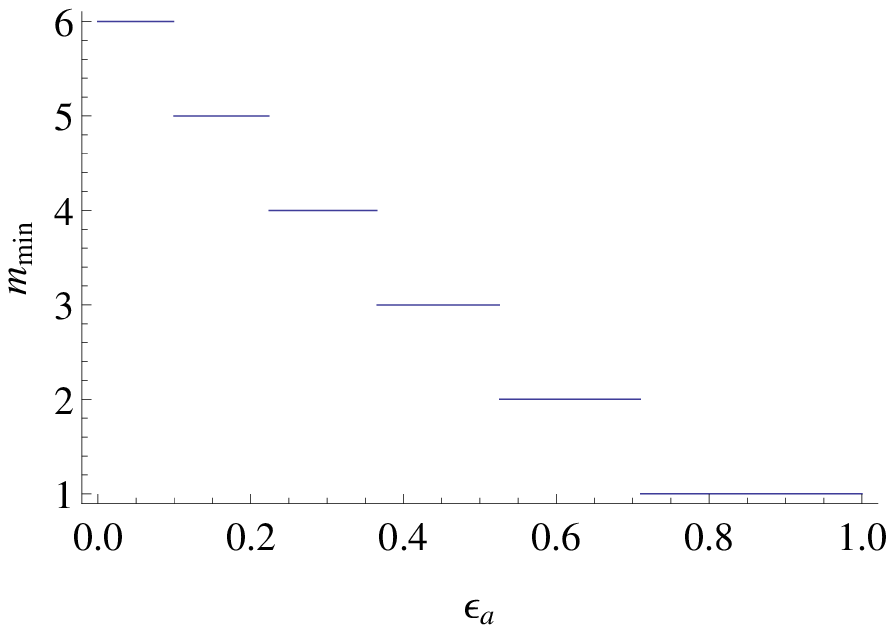}\\
  \caption{$m_{min}$ vs $\epsilon_a$ for $\epsilon_{g}$= $5.0 \times 10^{-5}$, using equations \ref{eq-mmin} only}
  \label{fig:mmin-ea}
\end{figure}

Figures \ref{fig:e_a_zero}, \ref{fig:e_a_0.3}, and \ref{fig:e_a_0.5} show the relationship between $P_L$ and $m$ for the case $\epsilon_{a}=0.0,0.3,0.5$, 
respectively, for gate error values $\epsilon_{g}=5.0 \times10^{-5},8.0 \times10^{-5},1.0 \times10^{-4}$, and $3.0 \times10^{-4}$. Note that $P_L$ is the cumulative error
for 1000 gates (plus QEC operations with some frequency), and hence may be larger than the underlying error for an individual physical gate, even when operating below threshold.
The black circular points are the values for  $P_L$ determined from the numerical simulation (section \ref{sec:level2}) and the the blue triangular points are for $P_L$ which are determined by using the exact
formula.

Our primary observations are that there is generally good agreement between the data generated by the formula and the simulation (and reasonably good agreement even
given the assumption of large $B$, especially with respect to the location of $m_{min}$), and that $m_{min}$ is insensitive to variations in $\epsilon_a$ over the
range of $\epsilon_a$ considered, with $m_{min}=5$ in all cases.  Figure \ref{fig:pl-ea} shows the behavior of $P_L$ as a function of $\epsilon_a$ for $\epsilon_g=5\times 10^{-5}$ and $m=5$, showing that the variation in $P_L$ is relatively small over lower values of $\epsilon_a$.  Figure \ref{fig:mmin-ea} shows the variation of the $m_{min}$, with $e_a$ for $\epsilon_g=5\times 10^{-5}$,
as given by equation \ref{eq-mmin}.

As expected, within the region $m<m_{min}$, the error rate is reduced both by increasing $m$ and by increasing $\epsilon_a$, since both result in fewer
QEC operations being performed (the only difference being whether the skipped operations are regularly spaced or not), and QEC operations
in this region produce more errors, on average, than they correct.  Similarly the behavior is reversed for $m>m_{min}$.

Overall agreement is good; there is a slightly larger disparity for $\epsilon_a=0$ and small $m$.  This is where the largest number of QEC operations occur
(since they are attempted frequently and none are skipped), indicating that the approximations in modeling QEC errors are leading to an
underestimate of the overall logical error.  At higher values of $m$ (where gate errors are a more dominant source of error) agreement is generally better,
except at the largest $m$ is large and the approximation is no longer valid.

\begin{table}
\caption{Table of possible second-order errors occurring within a single block.
Columns denote the source of the error (a set of $m$ gates $mG_i$,
a syndrome, double or omission error $Q_i(s,d,o)$ or a correction error $Q_i(c)$ during QEC)
and entries indicate the number of errors occurring in a particular location,
with a subscript (in the case of syndrome measurements) denoting the possible types.
The contribution column gives the total contribution to the logical error from all errors of that type
within $B$ blocks, thus the multiplier $B$.}
\label{table:errors-standard}
\centering
\begin{tabular}{|c| c| c| c|}
\hline
\multicolumn{3}{|c|}{Error source}&Error contribution\\ [0.5ex]
\hline
$mG_{i}$  & $Q_{i}(s,d,o)$  &$Q_{i}(c)$&(overall factor 42)         \\
\hline
2  	&    &   &$B(m\epsilon_g)^2/2$        \\
\hline
1  	&$1_{s,d}$    & &$B(1-\epsilon_a)m\epsilon_g(\epsilon_s+\epsilon_d)$        \\
\hline
        &$1_d+1_{s,d}$   & & $B(1-\epsilon_a)\epsilon_d(\epsilon_s+\epsilon_d/2)$      \\
\hline
        &$1_{s,o}$   &1    & $B(1-\epsilon_a)(\epsilon_s+\epsilon_o)\epsilon_c$      \\
\hline
  	&    &2    & $B(1-\epsilon_a)(\epsilon_c)^2/2$        \\
\hline
\end{tabular}
\end{table}

\begin{table*}
\caption{Table of possible second-order errors spanning 2 blocks.  Such errors can occur (even without
skipped QECs) when a QEC error in one block combines with an error (from logic gates or the QEC) in the subsequent block.  In
$B$ blocks there are $B-1$ sets of adjacent blocks, hence the contribution multiplier of $B-1$. 
}
\label{table:errors-2block}
\centering
\begin{tabular}{|c| c| c| c| c| c| c|}
\hline
\multicolumn{4}{|c|}{Error source}&Error contribution\\ [0.5ex]
\hline
$Q_{i}(s,d,o)$  &$Q_{i}(c)$&$mG_{i+1}$  & $Q_{i+1}(s)$&(overall factor 42)         \\
\hline
$1_{s,o}$    &    &1    &    &$(B-1)(1-\epsilon_a)(\epsilon_s+\epsilon_o) m\epsilon_g$        \\
\hline
$1_{s,o}$    &    &    &$1_{s,d}$    &$(B-1)(1-\epsilon_a)^2(\epsilon_s+\epsilon_o)(\epsilon_s+\epsilon_d)$\\ 	
\hline
  	    &1    &1    &    &$(B-1)(1-\epsilon_a)\epsilon_c m\epsilon_g$        \\
\hline
	    &1    &    &$1_{s,d}$    &$(B-1)(1-\epsilon_a)^2\epsilon_c(\epsilon_s+\epsilon_d)$		\\
\hline
\end{tabular}
\end{table*}

\begin{table}
\caption{Table of possible second-order errors due to $f$ skipped QECs spanning $f+1$ blocks.  This combines errors from operations
separated by one or more blocks with skipped QECs, where QECs from blocks $i$ to $i+f-1$ are skipped, and no logical gate errors occur
in blocks $i+1$ to $i+f-1$ (these operations, omitted from the table, are denoted by the column of Xs).  The contribution
includes contributions for all $f>0$, hence the multiplier $\gamma$, as discussed in section \ref{sec-logical}.
}
\label{table:errors-gamma}
\centering
\begin{tabular}{|c| c| c| c| c| c| c| c| c| c|}
\hline
\multicolumn{4}{|c|}{Error source}&Error contribution\\ [0.5ex]
\hline
$mG_{i}$  &X&$mG_{i+f}$& $Q_{i+f}(s,d,o)$&(overall factor 42)         \\
\hline
	1&X&   1   &    &$\gamma(m\epsilon_g)^2$		\\
\hline
	1&X&    &$1_{s,d}$   &$\gamma (1-\epsilon_a)m\epsilon_g(\epsilon_s+\epsilon_d)$		\\
\hline
\end{tabular}
\end{table}

\begin{table*}
\caption{Table of possible second-order errors due to $f$ skipped QECs, spanning $f+2$ blocks.  As with the analogous Tables \ref{table:errors-standard} and \ref{table:errors-2block},
error within QEC operations can lead to a total span of one additional block, compared to the case of Table \ref{table:errors-gamma}, leading to a multiplier $\gamma_3$, as discussed in
section \ref{sec-logical}.  QECs from blocks $i+1$ to $i+f$ are skipped, and no errors occur
in blocks $i+2$ to $i+f$.
}
\label{table:errors-gamma3}
\centering
\begin{tabular}{|c| c| c| c| c| c| c| c| c| c|}
\hline
\multicolumn{5}{|c|}{Error source}&Error contribution\\ [0.5ex]
\hline
$Q_{i}(s,d,o)$  &$Q_{i}(c)$&X&$mG_{i+1+f}$  & $Q_{i+1+f}(s)$  &(overall factor 42)         \\
\hline
	$1_{s,o}$   &    &X&1&& $\gamma_3 (1-\epsilon_a)m(\epsilon_s+\epsilon_o)\epsilon_g$		\\
\hline
	$1_{s,o}$   &    &X&&$1_{s,d}$& $\gamma_3 (1-\epsilon_a)^2(\epsilon_s+\epsilon_o)(\epsilon_s+\epsilon_d)$		\\	
\hline
	    &1    &X&1&& $\gamma_3 (1-\epsilon_a)\epsilon_cm\epsilon_g$		\\
\hline
	    &1    &X&&$1_{s,d}$& $\gamma_3 (1-\epsilon_a)^2\epsilon_c(\epsilon_s+\epsilon_d)$		\\	
\hline
\end{tabular}
\end{table*}

\section{Conclusion}

We have found the optimal number of quantum gates to perform before applying an error correction operation.  
This was done with a semi-abstract model to enable some generality.  An analytic expression was presented which provides explicit dependence on the error correction frequency as a function of the gate error rate, ancilla failure rate,
and error rates for the correction operation.  The various rates depend on the underlying physical gate error rate.  The dependence is different for different circuits which are determined by the code used.  To be explicit we showed in detail how this works by example.
Our example is the commonly used Steane [[7,1,3]] code and Steane ancilla technique.  However, we believe an immediate extension could be made to other distance-3 CSS codes, and a more general formula could be derived by a similar technique for other codes.  

We have assumed a transversal gate model.  Single-qubit logical operations may, however, depending on the
computation and code, be dominated by non-transveral gates (such as the $T$ gate).  Such gates, which often require preparation of a post-selected ancilla, would require an error model similar to that used for QEC operations.  Our treatment is expected to work very well for storage where examples include gates commonly called HOLD, or WAIT.  

For the Steane code and Steane ancilla, we compared the results with the detailed simulation using QASM-P.  We found excellent agreement showing that our course-graining (due to a rough classification of error types) provides enough precision to provide a very reliable estimate.   
Furthermore, we find that the optimum frequency to apply QEC operations is relatively insensitive to to ancilla failure
probability, over the range of gate error and ancilla error probabilities considered (with the optimum varying from $m=3$ to $m=6$ but frequency changes within this range making only small differences to the overall logical error), indicating that skipping QEC operations under ancilla failure will in many cases be a successful approach even in
 a design where QECs are performed infrequently.  This will save resources while providing a better overall logical error rate for quantum error correcting codes.  

\begin{acknowledgments}
The authors thank Brian Arnold, Ken Brown and Yaakov Weinstein for helpful discussions.
Supported by the Intelligence Advanced Research Projects Activity (IARPA) via Department of Interior National Business Center contract number D12PC00527. The U.S. Government is authorized to reproduce and distribute reprints for Governmental purposes notwithstanding any copyright annotation thereon. Disclaimer: The views and conclusions contained herein are those of the authors and should not be interpreted as necessarily representing the official policies or endorsements, either expressed or implied, of IARPA, DoI/NBC, or the U.S. Government.
\end{acknowledgments}

\bibliography{bib}

\begin{thebibliography}{12}%
\makeatletter
\providecommand \@ifxundefined [1]{%
 \@ifx{#1\undefined}
}%
\providecommand \@ifnum [1]{%
 \ifnum #1\expandafter \@firstoftwo
 \else \expandafter \@secondoftwo
 \fi
}%
\providecommand \@ifx [1]{%
 \ifx #1\expandafter \@firstoftwo
 \else \expandafter \@secondoftwo
 \fi
}%
\providecommand \natexlab [1]{#1}%
\providecommand \enquote  [1]{``#1''}%
\providecommand \bibnamefont  [1]{#1}%
\providecommand \bibfnamefont [1]{#1}%
\providecommand \citenamefont [1]{#1}%
\providecommand \href@noop [0]{\@secondoftwo}%
\providecommand \href [0]{\begingroup \@sanitize@url \@href}%
\providecommand \@href[1]{\@@startlink{#1}\@@href}%
\providecommand \@@href[1]{\endgroup#1\@@endlink}%
\providecommand \@sanitize@url [0]{\catcode `\\12\catcode `\$12\catcode
  `\&12\catcode `\#12\catcode `\^12\catcode `\_12\catcode `\%12\relax}%
\providecommand \@@startlink[1]{}%
\providecommand \@@endlink[0]{}%
\providecommand \url  [0]{\begingroup\@sanitize@url \@url }%
\providecommand \@url [1]{\endgroup\@href {#1}{\urlprefix }}%
\providecommand \urlprefix  [0]{URL }%
\providecommand \Eprint [0]{\href }%
\providecommand \doibase [0]{http://dx.doi.org/}%
\providecommand \selectlanguage [0]{\@gobble}%
\providecommand \bibinfo  [0]{\@secondoftwo}%
\providecommand \bibfield  [0]{\@secondoftwo}%
\providecommand \translation [1]{[#1]}%
\providecommand \BibitemOpen [0]{}%
\providecommand \bibitemStop [0]{}%
\providecommand \bibitemNoStop [0]{.\EOS\space}%
\providecommand \EOS [0]{\spacefactor3000\relax}%
\providecommand \BibitemShut  [1]{\csname bibitem#1\endcsname}%
\let\auto@bib@innerbib\@empty
\bibitem [{\citenamefont {Nielsen}\ and\ \citenamefont
  {Chuang}(2010)}]{nielsen}%
  \BibitemOpen
  \bibfield  {author} {\bibinfo {author} {\bibfnamefont {M.}~\bibnamefont
  {Nielsen}}\ and\ \bibinfo {author} {\bibfnamefont {I.}~\bibnamefont
  {Chuang}},\ }\href@noop {} {\emph {\bibinfo {title} {{Quantum Computation and
  Quantum Information}}}}\ (\bibinfo  {publisher} {Cambridge University Press,
  New York},\ \bibinfo {year} {2010})\BibitemShut {NoStop}%
\bibitem [{\citenamefont {Shor}(1997)}]{shor95}%
  \BibitemOpen
  \bibfield  {author} {\bibinfo {author} {\bibfnamefont {P.}~\bibnamefont
  {Shor}},\ }\href@noop {} {\bibfield  {journal} {\bibinfo  {journal} {Phys.
  Rev. A}\ }\textbf {\bibinfo {volume} {52}},\ \bibinfo {pages} {2493}
  (\bibinfo {year} {1997})}\BibitemShut {NoStop}%
\bibitem [{\citenamefont {Steane}(1996)}]{steane96}%
  \BibitemOpen
  \bibfield  {author} {\bibinfo {author} {\bibfnamefont {A.~M.}\ \bibnamefont
  {Steane}},\ }\href@noop {} {\bibfield  {journal} {\bibinfo  {journal} {Phys.
  Rev. Lett.}\ }\textbf {\bibinfo {volume} {77}},\ \bibinfo {pages} {793}
  (\bibinfo {year} {1996})}\BibitemShut {NoStop}%
\bibitem [{\citenamefont {Calderbank}\ and\ \citenamefont
  {Shor}(1996)}]{Calderbank96}%
  \BibitemOpen
  \bibfield  {author} {\bibinfo {author} {\bibfnamefont {A.}~\bibnamefont
  {Calderbank}}\ and\ \bibinfo {author} {\bibfnamefont {P.}~\bibnamefont
  {Shor}},\ }\href@noop {} {\bibfield  {journal} {\bibinfo  {journal} {Phys.
  Rev. A}\ }\textbf {\bibinfo {volume} {54}},\ \bibinfo {pages} {1098}
  (\bibinfo {year} {1996})}\BibitemShut {NoStop}%
\bibitem [{\citenamefont {Aharonov}\ and\ \citenamefont
  {Ben-Or}(2008)}]{aharonov08}%
  \BibitemOpen
  \bibfield  {author} {\bibinfo {author} {\bibfnamefont {D.}~\bibnamefont
  {Aharonov}}\ and\ \bibinfo {author} {\bibfnamefont {M.}~\bibnamefont
  {Ben-Or}},\ }\href@noop {} {\bibfield  {journal} {\bibinfo  {journal} {SIAM
  J. Comput.}\ }\textbf {\bibinfo {volume} {38}},\ \bibinfo {pages} {1207}
  (\bibinfo {year} {2008})}\BibitemShut {NoStop}%
\bibitem [{\citenamefont {Preskill}\ \emph {et~al.}(1998)\citenamefont
  {Preskill}, \citenamefont {Lo}, \citenamefont {Popescu},\ and\ \citenamefont
  {Spiller}}]{preskill98}%
  \BibitemOpen
  \bibfield  {author} {\bibinfo {author} {\bibfnamefont {J.}~\bibnamefont
  {Preskill}}, \bibinfo {author} {\bibfnamefont {H.-K.}\ \bibnamefont {Lo}},
  \bibinfo {author} {\bibfnamefont {S.}~\bibnamefont {Popescu}}, \ and\
  \bibinfo {author} {\bibfnamefont {T.~P.}\ \bibnamefont {Spiller}},\
  }\href@noop {} {\emph {\bibinfo {title} {Introduction to Quantum
  Computation}}}\ (\bibinfo  {publisher} {World Scientific, Singapore},\
  \bibinfo {year} {1998})\BibitemShut {NoStop}%
\bibitem [{\citenamefont {DiVincenzo}\ and\ \citenamefont
  {P.Aliferis}(2007)}]{div07}%
  \BibitemOpen
  \bibfield  {author} {\bibinfo {author} {\bibfnamefont {D.~P.}\ \bibnamefont
  {DiVincenzo}}\ and\ \bibinfo {author} {\bibnamefont {P.Aliferis}},\
  }\href@noop {} {\bibfield  {journal} {\bibinfo  {journal} {Phys. Rev. Lett.}\
  }\textbf {\bibinfo {volume} {98}},\ \bibinfo {pages} {020501} (\bibinfo
  {year} {2007})}\BibitemShut {NoStop}%
\bibitem [{\citenamefont {Abu-Nada}\ \emph {et~al.}(2014)\citenamefont
  {Abu-Nada}, \citenamefont {Fortescue},\ and\ \citenamefont {Byrd}}]{ali14}%
  \BibitemOpen
  \bibfield  {author} {\bibinfo {author} {\bibfnamefont {A.}~\bibnamefont
  {Abu-Nada}}, \bibinfo {author} {\bibfnamefont {B.}~\bibnamefont {Fortescue}},
  \ and\ \bibinfo {author} {\bibfnamefont {M.}~\bibnamefont {Byrd}},\
  }\href@noop {} {\bibfield  {journal} {\bibinfo  {journal} {Phys. Rev. A}\
  }\textbf {\bibinfo {volume} {89}},\ \bibinfo {pages} {062304} (\bibinfo
  {year} {2014})}\BibitemShut {NoStop}%
\bibitem [{\citenamefont {Weinstein}(2013)}]{yaa13}%
  \BibitemOpen
  \bibfield  {author} {\bibinfo {author} {\bibfnamefont {Y.~S.}\ \bibnamefont
  {Weinstein}},\ }\href@noop {} {\bibfield  {journal} {\bibinfo  {journal}
  {Phys. Rev. A}\ }\textbf {\bibinfo {volume} {88}},\ \bibinfo {pages} {012325}
  (\bibinfo {year} {2013})}\BibitemShut {NoStop}%
\bibitem [{\citenamefont {Weinstein}(2014)}]{yaa14}%
  \BibitemOpen
  \bibfield  {author} {\bibinfo {author} {\bibfnamefont {Y.~S.}\ \bibnamefont
  {Weinstein}},\ }\href {\doibase 10.1103/PhysRevA.89.020301} {\bibfield
  {journal} {\bibinfo  {journal} {Phys. Rev. A}\ }\textbf {\bibinfo {volume}
  {89}},\ \bibinfo {pages} {020301} (\bibinfo {year} {2014})}\BibitemShut
  {NoStop}%
\bibitem [{\citenamefont {Steane}(2003)}]{steane03}%
  \BibitemOpen
  \bibfield  {author} {\bibinfo {author} {\bibfnamefont {A.~M.}\ \bibnamefont
  {Steane}},\ }\href@noop {} {\bibfield  {journal} {\bibinfo  {journal} {Phys.
  Rev. A}\ }\textbf {\bibinfo {volume} {68}},\ \bibinfo {pages} {042322}
  (\bibinfo {year} {2003})}\BibitemShut {NoStop}%
\bibitem [{\citenamefont {Cross}(2006)}]{cross06}%
  \BibitemOpen
  \bibfield  {author} {\bibinfo {author} {\bibfnamefont {A.}~\bibnamefont
  {Cross}},\ }\href@noop {} {\enquote {\bibinfo {title} {{QASM}},}\ }\bibinfo
  {howpublished}
  {http://www.media.mit.edu/quanta/quanta-web/projects/qasm-tools} (\bibinfo
  {year} {2006})\BibitemShut {NoStop}%
\end{thebibliography}%
\end{document}